# IMPACTO DISTRIBUTIVO POTENCIAL DE REFORMAS NA TRIBUTAÇÃO INDIRETA NO BRASIL: SIMULAÇÕES BASEADAS NA PEC 45/2019

*Rozane Bezerra de Siqueira\**
*José Ricardo Bezerra Nogueira\**
*Carlos Feitosa Luna\*\*[1]*
(09 de abril de 2024)

**Abstract**
This paper analyzes the redistributive impacts of indirect taxation reforms in Brazil inspired by PEC 45/2019, particularly in the version that led to EC 132/2023. Comparisons are made between the current system and the simulated reforms, considering the distribution of the tax burden among families in different income classes, as well as the impact on poverty and inequality indicators. The simulations are conducted based on the combination of effective tax rates that apply to goods and services consumed by households, using nationally representative microdata from family budgets.

Resumo
Este artigo analisa os impactos redistributivos de reformas na tributação indireta no Brasil inspiradas na PEC 45/2019, em particular, na versão que deu origem à EC 132/2023. Comparações serão feitas do sistema corrente com as reformas simuladas, considerando a distribuição da carga tributária entre as famílias em diferentes classes de renda, bem como o impacto sobre indicadores de pobreza e desigualdade. A simulações são realizadas a partir da combinação das alíquotas tributárias efetivas que incidem sobre os bens e serviços consumidos pelas famílias com microdados nacionalmente representativos de orçamentos familiares.

**Palavras-chave**: reforma de tributos indiretos; alíquota padrão/modal, pobreza, desigualdade; microssimulação

**Códigos JEL**: H22, H23, C63

## 1 Introdução

Após mais de três décadas de discussão, em 20 de dezembro de 2023 foi promulgada a Emenda Constitucional n° 132 (EC 132/2023), oriunda da Proposta de Emenda Constitucional n°45 de 2019 (PEC 45/2019), que reforma a tributação do consumo no Brasil.

A reforma simplifica o sistema tributário atual ao substituir cinco tributos indiretos – ICMS, PIS, Cofins, IPI e ISS – por um Imposto sobre o Valor Agregado (IVA) dual, composto pela Contribuição sobre Bens e Serviços (IBS), da União, e pelo Imposto sobre Bens e Serviços (IBS), dos estados e municípios. Adicionalmente, a reforma cria um Imposto Seletivo (IS) que incidirá sobre bens e serviços que têm impacto negativo sobre a saúde e o meio ambiente.

---

[1] *Departamento de Economia, Universidade Federal de Pernambuco; \*\*Fundação Oswaldo Cruz/FIOCRUZ. Endereço para correspondência: rozane_siqueira@yahoo.com.br.

Na sua versão original a PEC 45/2019 propôs um IVA com alíquota única sobre todos os bens e serviços, com exceção para produtos que geram externalidades negativas, como bebidas alcóolicas e fumo. Para lidar com um possível efeito regressivo da reforma, a ideia era compensar as famílias de baixa renda via um esquema de devolução de imposto, o chamado "*cashback*".

Cedendo a pressões de grupos organizados que reivindicam tratamento tributário diferenciado, a versão aprovada da PEC 45 prevê vários regimes diferenciados (com isenções e alíquotas reduzidas) e específicos de tributação (em que a forma de cobrança do imposto pode ser diferente do modelo do IVA e alíquota pode ou não ser reduzida). Ao todo 29 itens ou categorias de bens e serviços foram elencados em um desses regimes especiais.

Isenções e alíquotas reduzidas são em geral defendidas com base em argumentos de equidade. No entanto, do nosso conhecimento, não há estudos avaliando a efetividade redistributiva da estrutura tributária prevista pela reforma aprovada. A importância dessa avaliação decorre de preocupações não apenas com a questão de equidade fiscal, mas também com os custos associados a alíquotas favorecidas, como a elevação da alíquota padrão do novo imposto e maior complexidade administrativa.

O objetivo deste artigo é analisar os impactos redistributivos de reformas na tributação indireta no Brasil inspiradas na PEC 45/2019, em particular, na versão que deu origem à EC 132/2023. Comparações serão feitas do sistema corrente com as reformas simuladas, considerando a distribuição da carga tributária entre as famílias em diferentes classes de renda, bem como o impacto sobre indicadores de pobreza e desigualdade. A simulações são realizadas a partir da combinação das alíquotas tributárias efetivas que incidem sobre os bens e serviços consumidos pelas famílias com microdados nacionalmente representativos de orçamentos familiares.

Cabe ressaltar que o detalhamento de quais itens de consumo exatamente serão favorecidos com isenção ou alíquota reduzida e sobre quais itens incidirá o Imposto Seletivo depende das leis complementares que ainda serão discutidas no Congresso Nacional. Esta indefinição impede uma avaliação acurada dos efeitos distributivos da reforma tributária aprovada e da alíquota padrão necessária para manter a arrecadação atual dos tributos que serão extintos. No entanto, exercícios de simulação de estruturas alternativas de alíquotas podem lançar alguma luz sobre essas questões.

O artigo está organizado em cinco seções. Depois desta introdução, a seção 2 descreve a estratégia empírica e os dados utilizados. A seção 3 examina o impacto distributivo do sistema corrente de tributos indiretos. A seção 4 define as reformas simuladas e investiga seus impactos redistributivos, comparando com o sistema tributário atual. Comentários finais são apresentados na seção 5.

## 2 Método e Dados

O cálculo da distribuição da carga tributária indireta entre as famílias requer três ingredientes: (i) as cestas de consumo das famílias, (ii) as alíquotas tributárias que incidem sobre essas cestas, e (iii) um indicador para avaliar o peso dos tributos sobre cada família, bem como para ordenar as famílias (ou os indivíduos) em termos de bem-estar.

Esta seção descreve os dados e procedimentos adotados neste estudo para realização desse cálculo. Ademais, a seção descreve brevemente os indicadores usados para mensurar o impacto da tributação do consumo sobre a desigualdade e a pobreza.

### 2.1 Dados de consumo das famílias

Neste trabalho, as cestas de consumo das famílias são obtidas a partir dos microdados da mais recente Pesquisa de Orçamentos Familiares – POF 2017-2018, conduzida pelo Instituto Brasileiro de Geografia e Estatística (IBGE, 2019).

A POF é uma pesquisa domiciliar, de natureza amostral, que inclui informação de 178.431 pessoas em 58.039 famílias, as quais, ponderadas pelos 'pesos amostrais' fornecidos pela pesquisa, são representativas da população brasileira. A POF contém informação de despesa monetária e não monetária de mais de oito mil itens de consumo das famílias, além de características demográficas e socioeconômicas, incluindo a renda disponível.

### 2.2 Alíquotas tributárias sobre o consumo das famílias

A tributação indireta vigente no Brasil consiste em um conjunto extremamente complexo de tributos que usam diferentes bases tributárias. A incidência efetiva desse sistema sobre o consumo das famílias foi estimada pioneiramente em Siqueira, Nogueira e Souza (2001), com uma atualização sendo realizada por Siqueira, Nogueira e Luna (2021).[2] Nesses estudos, as alíquotas tributárias efetivas sobre o consumo familiar são calculadas usando-se um modelo baseado na matriz de insumo-produto para o Brasil, o qual parte das receitas efetivamente arrecadadas pelo governo (líquidas de subsídios) e considera as relações intersetoriais, permitindo, assim, levar em conta os efeitos da tributação de insumos e a sonegação. A abordagem se baseia na hipótese de que os tributos são totalmente repassados para frente em cada etapa da cadeia de produção.[3]

---

[2] Outro estudo recente que estima alíquotas tributárias efetivas para o Brasil é Silveira, Palomo, Cornelio e Tolon (2022), também baseado na matriz de insumo-produto, com resultados em geral próximos dos obtidos por Siqueira, Nogueira e Luna (2021).

O presente estudo utiliza as alíquotas tributárias efetivas estimadas por Siqueira, Nogueira e Luna (2021) com base na Matriz de Insumo-Produto 2015, disponibilizada pelo IBGE em 2017 (para detalhes sobre a matriz, ver IBGE, 2018). Para tanto, é necessário realizar uma compatibilização dos agrupamentos de bens e serviços na matriz de insumo-produto com os agrupamentos na POF. Para facilitar esse procedimento, a incidência tributária final apresentada em Siqueira, Nogueira e Luna (2021) por setor de atividade foi desagregada por produto.[4] Dessa forma, foram obtidas 127 alíquotas efetivas para aplicação aos itens de consumo da POF.

Para dar uma ideia da magnitude das alíquotas da tributação indireta com que se deparam as famílias atualmente, a Tabela 1 apresenta as alíquotas efetivas médias calculadas em relação à despesa monetária das famílias na POF, tanto com os tributos 'por dentro' da base de cálculo, quanto 'por fora', para 15 grandes categorias de bens e serviços.

**Tabela 1:** Alíquotas tributárias efetivas médias com base na despesa monetária (%)

| Categorias de bens e serviços | Alíquotas efetivas | |
|---|---|---|
| | 'Por dentro' | 'Por fora' |
| Cesta básica de alimentos | 13,5 | 15,6 |
| Outros alimentos | 26,0 | 35,1 |
| Fumo e bebidas alcóolicas | 40,9 | 69,2 |
| Vestuário | 21,0 | 26,6 |
| Energia elétrica e gás | 33,8 | 51,1 |
| Aluguel | 5,6 | 5,9 |
| Bens e serviços do lar | 13,9 | 16,1 |
| Saúde | 15,3 | 18,1 |
| Transporte particular | 26,6 | 36,2 |
| Transporte público | 21,0 | 26,6 |
| Comunicação | 29,5 | 41,8 |
| Educação | 4,5 | 4,7 |
| Recreação e cultura | 24,7 | 32,8 |
| Higiene e cuidados pessoais | 21,8 | 27,9 |
| Outros bens e serviços | 10,7 | 12,0 |

Fonte: Cálculo dos autores a partir da Matriz de Insumo-Produto 2015 e POF 2017-2018.

---

[3] A hipótese de que os tributos sobre bens e serviços são totalmente repassados para o consumidor final através dos preços é padrão na literatura empírica. Todavia, em teoria, é possível que o repasse seja de menos de 100%, ou mesmo mais de 100%, dependendo da estrutura de cada mercado.

[4] A desagregação foi realizada simplesmente usando a participação da demanda pelo produto na demanda da atividade (a preços básicos). No cálculo da alíquota sobre bebidas alcóolicas foi utilizada também informação sobre as alíquotas nominais que incidem sobre os diversos tipos de bebidas.

Observa-se que no sistema tributário atual alimentação está sujeita a tributação substancial – mesmo a cesta básica de alimentos, sobre a qual se poderia esperar uma alíquota menor do que a estimada, devido a existência de isenções e alíquotas nominais reduzidas. Acontece que, no sistema vigente, uma grande parcela da tributação sobre alimentação provém da tributação de insumos usados na produção e distribuição de alimentos. Siqueira, Nogueira e Luna (2021) estima que essa parcela representa cerca 31% da tributação efetiva de alimentos. Vale notar que para a categoria 'outros alimentos' a alíquota média elevada deve-se também à presença das subcategorias 'alimentação fora do domicílio' e 'bebidas não-alcóolicas', esta última incluindo refrigerantes, que são tributadas mais pesadamente.

As categorias com as alíquotas efetivas mais baixas, além de aluguel, são aquelas onde há uma forte participação de serviços, a saber: 'bens e serviços do lar', 'saúde', 'educação' e 'outros bens e serviços'. No caso de 'bens e serviços do lar', destacam-se os 'serviços domésticos', que representam 22% da despesa na categoria e sobre os quais não incidem tributos indiretos. Isso mascara a presença na categoria de itens fortemente tributados, como 'eletrodomésticos', cuja alíquota efetiva 'por fora' é 44,2%. Quanto à baixa tributação de 'saúde' e 'educação', deve-se lembrar que, além dessas categorias estarem sujeitas a alíquotas nominais mais baixas devido a prevalência de serviços, são setores particularmente beneficiados por subvenções governamentais, sendo isso captado pelas alíquotas efetivas.

Com relação a 'recreação e cultura', apesar de serviços também terem participação importante nessa categoria, a alíquota efetiva média é relativamente elevada devido à presença de 'serviços de comunicação' (como internet e televisão por assinatura), que respondem por cerca de um quarto da despesa total na categoria e são fortemente tributados. Ressalve-se que as alíquotas efetivas apresentadas na Tabela 1 refletem a incidência tributária antes da Lei Complementar 194/2022, que reduziu a carga tributária sobre energia elétrica, combustíveis e comunicação.

A título de validação dessas estimativas, deve-se mencionar que a aplicação das alíquotas efetivas (desagregadas a nível de produto) aos microdados da POF resultou em uma carga tributária agregada equivalente a 97% da carga sobre as famílias estimada na Matriz de Insumo-Produto 2015 por Siqueira, Nogueira e Luna (2021). Dessa forma, nossa abordagem permite reproduzir satisfatoriamente o impacto relativo dos tributos indiretos sobre as famílias, apesar do valor absoluto da receita tributária ser subestimado. Isso acontece porque, como é sabido, a POF subestima o consumo das famílias em relação às Contas Nacionais. Segundo Ibarra, Rubião e Fleury (2021), que também se propõe a calcular a incidência dos tributos indiretos usando a POF 2017-2018, o consumo familiar agregado da POF corresponde a 80% do reportado nas Contas Nacionais.[5]

---

[5] O problema dos microdados de pesquisas de orçamentos familiares subestimar a receita dos tributos indiretos em relação às Contas Nacionais é bem conhecido na literatura. O'Donoghue, Baldini e Mantovani (2004) observa que para a grande maioria dos países para os quais o IVA é simulado a receita obtida fica

**2.3 Indicador de bem-estar**

De posse de informação sobre o consumo das famílias e sobre as alíquotas tributárias efetivas que incidem sobre esse consumo, pode-se calcular o montante de tributo pago por cada família na base de microdados. No entanto, a mensuração do impacto distributivo da tributação indireta é uma questão mais complexa, pois requer a escolha de um indicador – renda ou consumo – para ser usado como base no cálculo da carga tributária, bem como para ordenar as famílias em termos de padrão de vida ou bem-estar.

Inúmeros autores consideram a despesa de consumo como a variável mais apropriada para o cálculo da incidência dos tributos indiretos, uma vez que o uso da renda corrente tende a exagerar a regressividade desses tributos (ver, por exemplo, Thomas, 2020; Creedy, 1998; e Metcalf, 1994). Do ponto de vista teórico, isso acontece essencialmente porque a abordagem da renda ignora o fato de que o montante poupado no ano corrente – cuja participação na renda familiar tende a aumentar com renda – sofrerá a incidência da tributação indireta quando for revertido em consumo no futuro.

O motivo de ordem prática está relacionado à constatação de que os rendimentos das famílias que se encontram na base da distribuição de renda tendem a ser sub-reportados nas pesquisas domiciliares.[6] Isso pode estar associado ao fato de as famílias mais pobres serem beneficiárias de programas governamentais que têm como critério de elegibilidade a baixa renda, criando incentivos para subdeclarar rendimentos. Além disso, essas famílias frequentemente têm fluxos de rendimentos bastante irregulares, o que potencializa imprecisões na mensuração da 'renda regularmente recebida'.

Como alerta Gastaldi, Liberati, Pisano e Tedeschi (2017), a subdeclaração de rendimentos nas pesquisas domiciliares torna comum entre as famílias nos estratos mais baixos de renda situações em que o consumo é muito superior a renda, o que pode resultar em uma carga tributária irrealisticamente elevada quando medida em relação a renda.[7] Com o objetivo de reduzir anomalias, esses autores propõem um ajustamento nas rendas cujos valores são inferiores ao consumo. Para tanto, definem a diferença entre consumo $C$ e renda $Y$ como $\theta = C - Y \mid_{(C > Y)}$, e o montante de ajustamento como $g = z\theta$, onde z é a

---

entre 70% e 85% das estatísticas oficiais. As razões são múltiplas, além de elementos que levam à subestimação do consumo das famílias (como a subdeclaração do consumo de bebidas alcóolicas e fumo), há o fato de que o consumo familiar não é o único componente da demanda final por bens e serviços. Uma parte dos tributos recai sobre o consumo do governo, sobre instituições sem fins lucrativos e, no sistema brasileiro corrente, sobre exportações e investimento.

[6] Este fenômeno é observado mesmo em economias desenvolvidas. Meyer e Sullivan (2011, 2013), por exemplo, examinam dados para os Estados Unidos e encontram forte evidência de que o consumo é melhor mensurado nas pesquisas do que a renda para famílias mais pobres. Esses autores observam que os principais rendimentos sujeitos a subdeclaração são aqueles relativos ao trabalho por conta própria e a transferências (ou doações) públicas e privadas.

[7] No caso da POF, em aproximadamente 2% do total de domicílios, a renda disponível chega a ser inferior ao montante de tributos indiretos estimado a partir de suas cestas de consumo.

fração do déficit de renda que é preenchido. Assim, o ajuste proposto consiste em somar *g* à renda *Y*.

O Gráfico 1 ilustra, para o Brasil, a sensibilidade da incidência dos tributos indiretos a ajustamentos na renda disponível.[8] Pode-se observar que quando a carga tributária é calculada usando a renda reportada na POF sem nenhum ajustamento (z=0), o perfil de incidência mostra-se fortemente regressivo, mesmo com a exclusão dos domicílios com renda igual a zero. No entanto, fica evidente que a maior parte da regressividade observada depende da incidência no primeiro décimo de renda, onde a razão tributos/renda é extremamente elevada, enquanto entre o segundo décimo e o último, o declínio da carga tributária é bem mais suave. Note-se que os ajustamentos na renda afetam de forma acentuada apenas o perfil de incidência entre o primeiro e o segundo décimos, evidenciando que as irregularidades estão concentras na base da distribuição de renda.

**Gráfico 1**: Sensibilidade da incidência dos tributos indiretos a ajustamentos na renda disponível

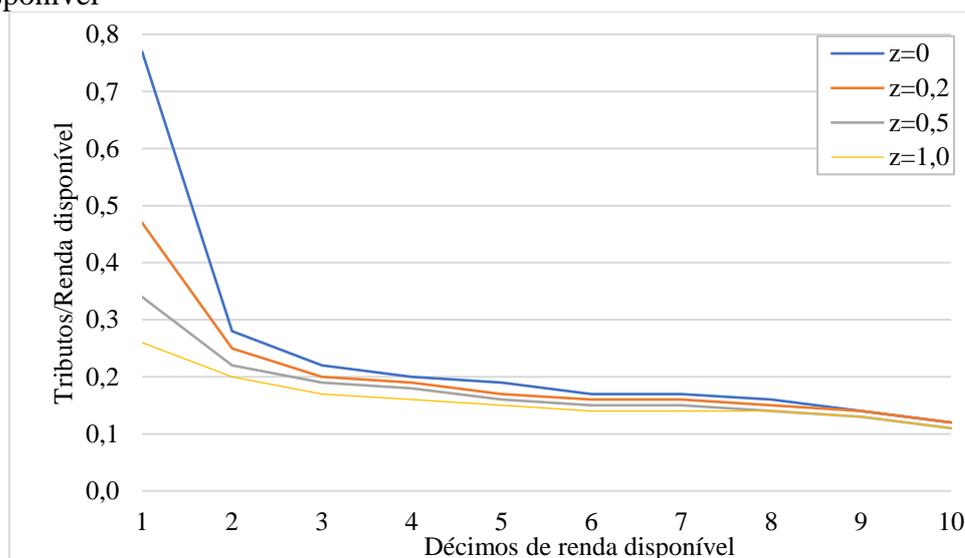

Fonte: Cálculo dos autores a partir da POF 2017-2018.

Levando isso em consideração, neste estudo, optamos por ajustar a renda fazendo z = 0,5, ou seja, aumentando a renda disponível monetária em metade do déficit observado em relação à despesa monetária de consumo. A opção pela renda disponível, em vez da renda bruta, tem o propósito de evitar a influência da tributação da renda nos resultados. Como alertam Thomas (2020) e Gastaldi, Liberati, Pisano e Tedeschi (2017), entre outros, o uso da renda bruta acentuaria a aparência regressiva da tributação indireta devido à progressividade da tributação da renda.

---

[8] Aqui, os ajustamentos partem da comparação da renda monetária disponível com a despesa monetária de consumo. A renda disponível é a renda bruta menos os tributos diretos (essencialmente, imposto de renda e contribuições previdenciárias). Na construção do Gráfico 1, para o caso em que z = 0, não foram consideradas as rendas iguais ou inferiores a 0.

Além da renda disponível ajustada, utilizamos a despesa total de consumo como base alternativa para calcular a distribuição da carga tributária entre as famílias. Em ambos os casos, visto que nosso objetivo é avaliar o impacto da tributação sobre o padrão de vida das famílias, as variáveis utilizadas compreendem montantes monetários e não monetários.

Para dar uma ideia do impacto sobre a renda das famílias do ajustamento realizado, a Tabela 2 compara a distribuição da renda disponível domiciliar *per capita* (incluindo a não-monetária) derivada diretamente da POF com a distribuição da renda ajustada (z = 0,5). Como esperado, a alteração mais forte em termos proporcionais ocorre no primeiro percentil da distribuição da renda ajustada, onde o ponto de corte (separatriz) é 38% mais elevado do que na distribuição da renda não-ajustada. Ao longo da distribuição, a diferença percentual entre os pontos de corte decresce, caindo para 12% já no décimo percentil e para 2% no último percentil.

**Tabela 2**: Efeito de ajustar as rendas

| | Renda disponível não ajustada | Renda disponível ajustada |
|---|---|---|
| Percentis | | |
| 1% | 126 | 173 |
| 5% | 270 | 320 |
| 10% | 392 | 441 |
| 25% | 700 | 751 |
| 50% | 1.238 | 1.296 |
| 75% | 2.131 | 2.260 |
| 90% | 3.895 | 4.077 |
| 95% | 5.899 | 6.164 |
| 99% | 13.152 | 13.467 |
| Obs. | 68.862.296 | 68.862.296 |
| Média | 1.957,72 | 2.048,75 |
| Desvio | 3.069,66 | 3.104,64 |

Fonte: Cálculo dos autores com base na POF 2017-2018.

### 2.4 Indicadores de pobreza e desigualdade

Complementando a análise da incidência da carga tributária por estrato de renda e despesa, este trabalho também calcula o impacto dos tributos indiretos sobre indicadores de desigualdade e pobreza.

Para examinar o efeito da tributação indireta sobre os pobres são usados índices da família Foster-Greer-Thorbecke (1984) de medidas de pobreza, que seguem a forma:

$$P_\alpha = \frac{1}{n}\sum_{i=1}^{q}\left(\frac{z-y_i}{z}\right)^\alpha$$

em que α é um parâmetro que mede a sensibilidade do índice à intensidade da pobreza (medida pela diferença entre a renda (ou despesa) do pobre e a linha de pobreza), z é a linha de pobreza, $y_i$ é o indicador de bem-estar escolhido (renda ou consumo), n é o número de indivíduos na população, e q é o número de pobres.

Apresentaremos resultados para três valores de α: 0,1, e 2. Note-se que para α = 0 o índice mede a proporção de pobres na população. Com α = 1, o índice representa o 'hiato de pobreza' – a distância média do indicador de bem-estar (renda ou consumo) dos pobres em relação à linha de pobreza (como proporção da linha de pobreza). Por fim, com α = 2, obtém-se o 'hiato quadrático', que dá um peso maior à insuficiência de renda (ou consumo) daqueles que estão mais distantes da linha de pobreza.

A linha de pobreza adotada neste estudo é a linha proposta pelo Banco Mundial para países de renda média alta: US$ 5,50 por dia. Em 2018, ano de referência para este estudo, essa linha era equivalente a R$ 420,00 por pessoa por mês – convertidas pela paridade de poder de compra (PPC 2011).

O indicador de desigualdade utilizado é o bem conhecido coeficiente de Gini. Este coeficiente varia de 0 a 1, sendo que quanto mais próximo de 1, maior a desigualdade. O impacto distributivo de uma reforma pode ser medido como a diferença entre os coeficientes de Gini antes da reforma e o Gini pós-reforma.

Tanto a pobreza quanto a desigualdade são aqui mensuradas com base em ambos os indicadores de bem-estar: a renda disponível domiciliar ajustada *per capita* e a despesa total de consumo domiciliar *per capita*.

## 3  Efeito Distributivo da Tributação Indireta Vigente

Para melhor interpretar os efeitos das reformas tributárias simuladas na próxima seção, é importante ter uma boa compreensão do sistema atual de tributos indiretos. Esta seção é dedicada a examinar o perfil distributivo da tributação indireta vigente, considerando a distribuição da carga entre as famílias em diferentes estratos de renda e impactos sobre a pobreza e a desigualdade no Brasil.

### 3.1 Distribuição da carga tributária entre as famílias

Para investigar a distribuição do peso da tributação do consumo, classificamos as famílias (ou domicílios) em décimos.[9] Os décimos são construídos ordenando todos os domicílios da POF de acordo com a renda disponível (ajustada) domiciliar *per capita* e dividindo-os em dez grupos, de forma que o primeiro décimo contém os 10% 'mais pobres' dos domicílios do país e o último décimo contém os 10% 'mais ricos'.

---

[9] Neste artigo usamos o termo família e domicílio de forma intercambiável. Ressalve-se, no entanto, que, de acordo com as definições da POF, a unidade de análise aqui utilizada é o domicílio.

O efeito distributivo da tributação indireta depende da combinação da diferenciação das cestas de consumo das famílias entre as classes de renda e das alíquotas tributárias efetivas que recaem sobre essas cestas. Assim, antes de analisar a incidência dos tributos indiretos e seus efeitos distributivos, é útil visualizar a estrutura de consumo das famílias em cada décimo de renda. A Tabela 3 mostra as parcelas orçamentárias médias das famílias por décimo de renda, para as 15 categorias de bens e serviços já apresentas na seção 2.[10] Note-se que essas parcelas foram calculadas em relação à despesa total de consumo (incluindo a despesa não-monetária).

**Tabela 3:** Parcelas orçamentárias médias por décimo de renda (%)

| Categorias de despesa | 1 | 2 | 3 | 4 | 5 | 6 | 7 | 8 | 9 | 10 | Todos |
|---|---|---|---|---|---|---|---|---|---|---|---|
| Cesta básica de alimentos | 14,4 | 12,0 | 10,5 | 9,7 | 9,6 | 8,3 | 7,9 | 6,8 | 5,7 | 4,0 | 8,9 |
| Outros alimentos | 8,1 | 8,6 | 8,7 | 9,1 | 8,7 | 8,6 | 9,3 | 9,2 | 9,4 | 8,8 | 8,9 |
| Bebidas e fumo | 0,8 | 0,9 | 0,9 | 1,0 | 0,9 | 1,0 | 1,0 | 1,0 | 0,9 | 0,7 | 0,9 |
| Vestuário | 4,7 | 4,8 | 4,4 | 4,5 | 4,2 | 4,1 | 3,8 | 3,6 | 3,5 | 3,0 | 4,1 |
| Eletricidade e gás | 10,5 | 8,7 | 7,9 | 7,1 | 6,9 | 6,5 | 5,6 | 5,1 | 4,1 | 2,9 | 6,5 |
| Aluguel | 22,6 | 22,3 | 22,8 | 22,1 | 22,4 | 24,0 | 22,7 | 22,7 | 21,6 | 20,0 | 22,3 |
| Bens e serviços do lar | 8,0 | 8,0 | 8,0 | 7,8 | 8,3 | 7,9 | 7,6 | 8,2 | 9,1 | 11,9 | 8,5 |
| Saúde | 5,2 | 5,7 | 6,8 | 6,9 | 7,7 | 7,9 | 7,9 | 7,6 | 8,7 | 8,9 | 7,3 |
| Transporte particular | 4,7 | 6,2 | 7,4 | 8,7 | 8,5 | 9,3 | 10,7 | 12,4 | 13,3 | 14,7 | 9,6 |
| Transporte público | 4,1 | 4,0 | 3,9 | 3,8 | 3,9 | 3,5 | 3,3 | 3,0 | 2,8 | 2,7 | 3,5 |
| Comunicação | 2,4 | 3,3 | 3,6 | 3,7 | 3,8 | 3,7 | 4,0 | 3,8 | 3,9 | 3,7 | 3,6 |
| Educação | 2,1 | 2,2 | 2,1 | 2,3 | 2,3 | 2,3 | 2,7 | 3,0 | 3,1 | 3,5 | 2,6 |
| Recreação e cultura | 3,3 | 3,9 | 4,0 | 4,1 | 3,7 | 3,9 | 4,1 | 4,2 | 4,4 | 4,8 | 4,0 |
| Higiene e cuidados pessoais | 7,5 | 7,2 | 6,5 | 6,3 | 5,7 | 5,5 | 5,2 | 4,8 | 4,2 | 3,4 | 5,6 |
| Outros bens e serviços | 1,6 | 2,2 | 2,6 | 2,9 | 3,4 | 3,6 | 4,3 | 4,5 | 5,3 | 6,9 | 3,7 |

Fonte: Cálculo dos autores com base na POF 2017-2018.

Destacam-se nessa tabela três categorias de bens e serviços cuja participação no orçamento é bastante elevada para famílias na base da distribuição de renda, mas cai fortemente à medida que a renda aumenta, a saber: 'cesta básica de alimentos', 'energia elétrica e gás' e 'higiene e cuidados pessoais'. Observe que, no sistema corrente, a alíquota efetiva sobre a cesta básica é reduzida em relação à média global, porém, as alíquotas sobre 'higiene e cuidados pessoais' e sobre 'energia elétrica e gás' são bem elevadas (ver Tabela 1).

---

[10] Na Tabela 3, e em todas as outras tabelas deste estudo onde são apresentadas estimativas por classe de renda, as médias apresentadas são calculadas a partir dos microdados, ou seja, a partir dos valores estimados para cada domicílio na POF – não são, portanto, 'médias agregadas', que seriam obtidas, por exemplo, dividindo-se a despesa agregada do décimo de renda em dado item pela despesa total agregada do décimo.

Por outro lado, as parcelas orçamentárias para 'saúde', 'educação' e 'outros bens e serviços' aumentam ao longo da distribuição de renda, ao mesmo tempo que suas alíquotas tributárias efetivas estão bem abaixo da média. Destaca-se ainda o forte aumento da participação de 'transporte privado' no orçamento familiar à medida que a renda aumenta, sendo que sobre esse item a alíquota efetiva estimada é uma das mais elevadas.

O resultado da combinação das alíquotas tributárias efetivas com a estrutura de consumo das famílias em termos de distribuição da carga tributária é apresentado na Tabela 4. Essa tabela apresenta as cargas tributárias médias com as quais se deparam as famílias como percentagem da renda disponível e como percentagem da despesa total de consumo, por décimos de renda.

**Tabela 4:** Carga tributária média (%)

| Décimos de renda disponível ajustada | Despesa de consumo | Renda disponível ajustada ($z = 0{,}5$) |
|---|---|---|
| 1 | 15,7 | 15,9 |
| 2 | 15,7 | 14,4 |
| 3 | 15,6 | 13,6 |
| 4 | 15,7 | 13,5 |
| 5 | 15,4 | 12,5 |
| 6 | 15,1 | 12,3 |
| 7 | 15,2 | 12,4 |
| 8 | 15,1 | 12,4 |
| 9 | 15,0 | 12,0 |
| 10 | 14,5 | 10,6 |
| **Todos os domicílios** | **15,3** | **13,0** |

Fonte: Cálculo dos autores com base na POF 2017-2018.

Pode-se observar que, calculada com base na despesa total de consumo (incluindo a não-monetária), a incidência da tributação de bens e serviços no Brasil é aproximadamente proporcional, porém, com um leve viés de regressividade, com o peso dos tributos caindo de 15,7% no orçamento no primeiro décimo para 14,5% no último décimo.

A quase proporcionalidade da incidência dos tributos indiretos quando calculada em relação à despesa total é também observada em Silveira, Palomo, Cornelio e Tonon (2022), que estima uma carga tributária variando levemente em torno de 15% entre os décimos. Cabe mencionar que, estimando a carga tributária com relação apenas ao componente monetário da despesa de consumo, tanto Silveira, Palomo, Cornelio e Tonon

(2022) quanto Ibarra, Rubião e Fleury (2021) encontram uma incidência levemente regressiva para a tributação do consumo no Brasil.[11]

Ainda que pouco significativa, a regressividade observada na incidência dos tributos indiretos no Brasil em relação à despesa de consumo é chocante, pois significa que, apesar das isenções e multiplicidade de alíquotas, frequentemente defendidas com base em argumentos de equidade, o sistema vigente fracassa em gerar um perfil minimamente pró-pobre. Em contraste, Thomas (2020) encontra que na maioria dos países OCDE, onde a tributação do consumo é bem menos complexa e diferenciada por setor ou produto do que no Brasil atualmente, a distribuição da carga tributária entre as famílias ou é proporcional ou levemente progressiva quando medida como uma percentagem da despesa de consumo.

De fato, esse viés regressivo do sistema brasileiro já poderia ser previsto a partir da análise das alíquotas tributárias efetivas e das cestas de consumo das famílias em diferentes estratos de renda. Uma incidência progressiva só surgiria se a estrutura de alíquotas fosse tal que sistematicamente alíquotas mais baixas fossem impostas sobre bens e serviços que têm um peso mais elevado no orçamento das classes de renda mais baixa. Como vimos, no sistema brasileiro atual não é isso que se observa, em grande medida devido à subtributação de serviços.

Estimada em relação à renda disponível ajustada (incluindo a renda não-monetária), a incidência dos tributos indiretos se mostra mais regressiva, com a carga tributária média variando de 15,9%, sobre o primeiro décimo de renda, a 10,6%, sobre o último décimo. Vale observar que outros autores encontram um grau de regressividade bem mais elevado do que este estudo quando a base de cálculo é a renda, em parte por não tratarem dos déficits orçamentários observados na base da distribuição de renda (caso de Silveira, Palomo, Cornelio e Tonon, 2022) ou por considerar apenas a renda monetária (caso de Ibarra, Rubião e Fleury, 2021). Além disso, esses autores usam a renda bruta (em vez da renda disponível) como base de cálculo da carga tributária.

### 3.2 Impacto sobre a desigualdade e a pobreza

Esta subseção complementa a análise distributiva do sistema vigente de tributos indiretos calculando seu impacto sobre a desigualdade, medida pelo coeficiente de Gini, e sobre diferentes indicadores de pobreza. As Tabelas 5 e 6 apresentam os índices de pobreza e desigualdade com base na despesa total de consumo domiciliar *per capita* e na renda disponível domiciliar *per capita*, respectivamente. Em cada caso, as medidas de pobreza e desigualdade são calculadas, primeiro, para o valor bruto do indicador de bem-estar, isto é, sem deduzir os tributos indiretos, e, depois, subtraindo os tributos, obtendo-se assim a 'despesa líquida' e a 'renda líquida'.[12]

---

[11] Ressalve-se que há importantes diferenças metodológicas entre este trabalho e Ibarra, Rubião e Fleury (2021), em particular, o fato desses autores usarem em suas estimativas os tributos cobrados pelo estado de São Paulo como representativos do Brasil como um todo.

**Tabela 5:** Impacto sobre a pobreza e a desigualdade com base na despesa de consumo

| Indicadores | Despesa bruta | Despesa líquida | Variação |
|---|---|---|---|
| Proporção de pobres | 0,208 | 0,272 | 30,7% |
| Hiato de pobreza | 0,069 | 0,096 | 38,6% |
| Hiato quadrático | 0,033 | 0,047 | 43,7% |
| Coeficiente de Gini | 0,491 | 0,493 | 0,4% |

Fonte: Cálculo dos autores com base na POF 2017-2018.

A Tabela 5 mostra que, com base na despesa total de consumo (incluindo a despesa não-monetária), 20,8% da população do Brasil em 2018 era pobre, ou seja, era composta de indivíduos que moravam em domicílios cuja despesa de consumo *per capita* era inferior à linha de pobreza de R$ 420 por mês, por pessoa. Após a dedução dos tributos indiretos, esse percentual sobe para 27,2%, um aumento de 30,7%. A intensidade da pobreza, medida pelo hiato de pobreza, tem um aumento ainda mais significativo, principalmente quando mensurada pelo hiato quadrático, indicando que os mais pobres entre os pobres são mais penalizados pela tributação do consumo. Por sua vez, o aumento na desigualdade na distribuição da despesa de consumo, medida pelo coeficiente de Gini, é muito modesto, 0,4%, confirmando os resultados da análise por décimos de renda, que mostram um sistema muito próximo da proporcionalidade, mas com um leve viés de regressividade.

A Tabela 6 mostra que, estimada com base na renda disponível, a proporção de pobres no Brasil em 2018, antes de deduzidos os tributos sobre o consumo, era 13,4%. A subtração dos tributos indiretos da renda aumenta em cerca de um terço esse percentual. O hiato de pobreza também sofre um aumento relativamente maior na Tabela 6 do que na Tabela 5. Isso acontece porque quando a renda disponível é utilizada, em vez da despesa de consumo, os mais pobres ficam mais próximos da linha de pobreza e, portanto, variações na renda tendem a ter um impacto maior sobre os indicadores de pobreza. Como esperado, o impacto dos tributos indiretos sobre o coeficiente de Gini também é mais significativo quando medido em relação à renda do que ao consumo, refletindo as distribuições da carga tributária apresentadas na Tabela 4.

**Tabela 6:** Impacto sobre a pobreza e a desigualdade com base na renda disponível

| Indicadores | Renda disponível bruta | Renda disponível líquida | Variação |
|---|---|---|---|
| Proporção de pobres | 0,134 | 0,179 | 33,7% |
| Hiato de pobreza | 0,042 | 0,060 | 43,5% |
| Hiato quadrático | 0,019 | 0,029 | 51,1% |
| Coeficiente de Gini | 0,496 | 0,506 | 1,9% |

Fonte: Cálculo dos autores com base na POF 2017-2018.

---

[12] Frequentemente a literatura se refere à renda disponível líquida de tributos indiretos como 'renda consumível' ou renda 'pós-fiscal', por exemplo, Lustig (2018).

# 4 IMPACTO DISTRIBUTIVO DE REFORMAS BASEADAS NA PEC 45/2019

Esta seção investiga os efeitos distributivos potenciais de reformas na tributação indireta no Brasil inspiradas na PEC 45/2019, tanto na sua versão inicial quanto na versão que deu origem à EC 132/2023. Além de examinarmos as implicações redistributivas da diferenciação de alíquotas, avaliamos também seu custo em termos do impacto sobre a alíquota padrão. Antes, porém, é útil descrevermos brevemente a estrutura de alíquota prevista em cada versão da PEC 25.

## 4.1 Estruturas de tributação na PEC 45/2019

Como já mencionado, inicialmente a PEC 45/2019 previu um IVA uniforme sobre todos os bens e serviços, com exceção para produtos prejudiciais à saúde e ao meio ambiente, como bebidas alcóolicas e fumo. Para garantir que a reforma não fosse regressiva, a proposta original era compensar os contribuintes de baixa renda via um mecanismo de devolução do imposto – frequentemente chamado de "*cashback*".

No entanto, durante a tramitação da PEC 45 no Congresso Nacional a ideia de alíquota única não prevaleceu. O texto aprovado, que resultou na EC 132/2023, prevê quatro alíquotas, além de um imposto seletivo sobre bens e serviços nocivos à saúde e ao meio ambiente, a saber: a) uma alíquota padrão; b) alíquota zero sobre uma cesta básica de alimentos a ser definida em lei complementar; c) alíquota de 40% da alíquota padrão para determinados itens de consumo; e d) alíquota de 70% da alíquota padrão para serviços de profissão intelectual, de natureza científica, literária ou artística.

As categorias de bens e serviços favorecidas com a redução de 60% da alíquota padrão são: (i) serviços de educação; (ii) serviços de saúde; (iii) dispositivos médicos; (iv) dispositivos de acessibilidade para pessoas com deficiência; (v) medicamentos; (vi) produtos de cuidados básicos à saúde menstrual; (vii) serviços de transporte público coletivo de passageiros rodoviário e metroviário de caráter urbano; semiurbano e metropolitano; (viii) alimentos destinados ao consumo humano; (ix) produtos básicos de higiene pessoal e limpeza majoritariamente consumidos por famílias de baixa renda; (x) produtos agropecuários, aquícolas, pesqueiros, florestais e extrativistas vegetais *in natura*; (xi) insumos agropecuários e aquícolas; (xii) produções artísticas, culturais, de eventos, jornalísticas e audiovisuais nacionais, atividades desportivas e comunicação institucional; e (xiii) bens e serviços associados à soberania nacional, segurança da informação e cibernética. Ressalte-se que para alguns subitens dessa lista, a serem definidos na legislação complementar, a EC 132/2023 prevê a possibilidade de isenção ou alíquota zero.

A EC 132/2023 também prevê regimes específicos de tributação – ou seja, regimes diferenciados em relação ao modelo de cobrança do IVA – para alguns bens e serviços, com a possibilidade de alíquota favorecida em relação à alíquota padrão. Os seguintes itens são elencados como sujeitos a regimes específicos: combustíveis e lubrificantes;

serviços financeiros; planos de assistência à saúde; operações com bens imóveis; concursos e prognósticos; operações de sociedades cooperativas; serviços de hotelaria, parques de diversão e parques temáticos, agências de viagem e de turismo, bares e restaurantes, sociedade anônima de futebol e aviação regional; e serviços de transporte coletivo de passageiros rodoviário intermunicipal e interestadual, ferroviário e hidroviário.

A EC 132/2023 cria também a possibilidade de *cashback* para a população de baixa renda com o objetivo compensar, em alguma medida, o imposto incidente sobre energia elétrica e gás de cozinha.

Por fim, cabe mencionar que a reforma aprovada estabelece que as alíquotas do novo imposto serão fixadas de modo a manter a carga tributária atual dos cinco tributos que serão extintos.

### 4.2 Reformas simuladas

Embora o foco da nossa análise seja nas reformas baseadas na PEC 45/2019, duas outras simulações são realizadas com o objetivo de melhorar a compreensão das implicações distributivas da diferenciação de alíquotas, tanto no sistema tributário atual e quanto no arcabouço da PEC 45. Dessa maneira, quatro reformas são simuladas.

A **Reforma 1** supõe um IVA com alíquota uniforme única sobre todos os bens e serviços, sem exceção. A **Reforma 2** é baseada na primeira versão da PEC 45/2019 e também propõe um IVA com alíquota uniforme, mas admite a presença de um imposto seletivo sobre fumo e bebidas alcóolicas, bem como a devolução para os domicílios pobres de todo imposto pago sobre o consumo de bens e serviços. **A Reforma 3** procura captar a estrutura de alíquotas prevista na EC 132/2023. Por fim, a **Reforma 4** consiste no exercício de substituir a isenção da cesta básica de alimentos, no arcabouço da Reforma 3, por um montante fixo de dinheiro pago a cada indivíduo na população.

Em todas as simulações a alíquota padrão é determinada de forma que a carga tributária global sobre consumo das famílias, líquida de *cashback*, seja a mesma do sistema corrente – conforme estimada na seção 3. Em outras palavras, as reformas simuladas são neutras do ponto de vista orçamentário, não gerando déficits ou superávits.

Por fim, vale esclarecer que as alíquotas tributárias apresentadas nesta seção são estimadas com base na despesa monetária de consumo e, como prevê a reforma tributária aprovada, são calculadas com o imposto 'por fora'.[13]

A seguir descrevemos em mais detalhes as quatro reformas simuladas, já informando a alíquota padrão estimada para cada uma delas. Novamente, vale ressaltar que as estruturas tributárias simuladas com base na PEC 45 não são representações precisas da proposta, em parte, devido a dificuldades técnicas de modelar todos os detalhes da reforma aprovada e, em parte, porque as definições de quais itens exatamente estarão sujeitos `as alíquotas favorecidas e ao imposto seletivo, bem como a forma de tributação nos regimes específicos, dependem ainda de lei complementar.

---

[13] Ressalte-se que serviços domésticos são excluídos da base de cálculo da carga tributária.

Reforma 1:
- **Alíquota uniforme de 25,8%** sobre todos os bens e serviços, sem exceção.

Reforma 2:
- **Alíquota padrão de 26,7%** sobre todos os bens e serviços, exceto 'fumo e bebidas alcoólicas'.
- **Imposto Seletivo sobre 'fumo e bebidas alcóolicas'** com alíquota igual a duas vezes a alíquota-padrão.
- *Cashback* de todo imposto sobre o consumo de bens e serviços, exceto 'fumo e bebidas alcóolicas', para os domicílios cuja despesa total domiciliar *per capita* é inferior à linha de pobreza de R$ 420 por mês por pessoa.

Reforma 3:
- **Alíquota zero** para produtos da **cesta básica** de alimentos.[14]
- **Alíquota padrão de 33,2%**.
- **Alíquota 40% da alíquota padrão** sobre: serviços de educação, serviços de saúde, medicamentos, transporte público urbano, semiurbano e metropolitano, produções artísticas, culturais e atividades desportivas.
- **Alíquota 70% da alíquota padrão** sobre: serviços prestados por profissionais liberais[15]
- **Alíquota sobre serviços financeiros e planos de assistência à saúde** mantida igual à alíquota efetiva média no sistema vigente, estimada em 21,4% ('por fora').
- **Imposto Seletivo sobre 'fumo e bebidas alcóolicas'** com alíquota igual a duas vezes a alíquota-padrão.
- **Cashback do imposto pago sobre energia elétrica e gás** para os domicílios cuja despesa total domiciliar *per capita* é inferior à linha de pobreza de R$ 420 por mês por pessoa.

Reforma 4
- Igual à Reforma 3, porém, com a cesta básica sendo tributada à alíquota padrão e a receita extra sendo usada para financiar **uma transferência de montante fixo para todos** os indivíduos na população.

---

[14] Neste estudo, a cesta básica é composta de arroz, feijão, café em pó, açúcar, leite de vaca e em pó integral, pescados, carnes, ovos, farinha de trigo, farinha de mandioca, pão francês, óleo de soja, manteiga e margarina, tubérculos e raízes, legumes e verduras, frutas, outros cereais, leguminosas e oleaginosas.

[15] Serviços jurídicos respondem por mais de 80% do consumo sujeito a essa alíquota.

Antes de analisar os impactos redistributivos dessas reformas, cabe considerar em que medida as alíquotas efetivas e a carga tributária global estimadas para o sistema corrente de tributos indiretos podem servir de referência para as simulações da PEC 45/2019. Conforme já destacado, essas alíquotas refletem a incidência final sobre as famílias do total de tributos indiretos líquido de subsídios e foram estimadas a partir da matriz de insumo-produto. Sendo assim, um ponto a observar é que, do total de tributos sobre produtos computados na matriz, os cinco tributos alvo da reforma (ICMS, Cofins, PIS, IPI e ISS) representam 89%. Portanto, é razoável supor que esses cinco tributos determinam a atual estrutura de alíquotas, em termos de grau de diferenciação.

Para avaliar a adequação do nível da carga tributária estimada deve-se também levar em conta que, no sistema corrente, do total de tributos indiretos, apenas 76% têm incidência final sobre o consumo das famílias, com a parcela restante recaindo sobre investimentos, exportações e o consumo do governo (ver Siqueira, Nogueira e Luna, 2021). Foi observado ainda que, o montante total de tributos indiretos com incidência final sobre o consumo das famílias, na matriz de insumo-produto, é equivalente a 86% da receita dos cinco tributos unificados pela reforma. Assim, com base nesses cálculos, uma vez que a reforma se propõe a desonerar o investimento e as exportações sem reduzir a carga tributária, há a possibilidade da alíquota-padrão sobre as famílias em nossas simulações estar subestimada.

Porém, também é importante levar em consideração que o valor da alíquota padrão depende da forma como os regimes diferenciados e específicos são modelados. Observe-se que em nossas simulações, enquanto os itens incluídos nos regimes diferenciados o foram de forma irrestrita (por exemplo, incluímos todos os medicamentos e serviços de saúde), alguns itens também sujeitos a alíquota reduzida segundo a reforma aprovada – como produtos básicos de limpeza e higiene pessoal – não foram tratados de forma diferenciada nas simulações. Ademais, as simulações não consideram os regimes específicos, exceto para serviços financeiros e planos de saúde.

Apesar de diferenças metodológicas marcantes, bem como em relação ao desenho da reforma simulada, pode ser interessante comparar nossas estimativas da alíquota padrão com estudos anteriores.

Orair e Gobetti (2019) foi pioneiro em oferecer uma estimativa da alíquota neutra do ponto de vista arrecadatório para um IVA uniforme no contexto da versão inicial da PEC45/2019, obtendo uma alíquota de 26,9%. Mais recentemente, o Instituto Mauro Borges (IMB, 2023) estimou que essa alíquota poderia alcançar o valor máximo de 23,4%, dependendo das premissas de cálculo. Por sua vez, um estudo do Ministério da Fazenda (2023) estimou que, na presença do Imposto Seletivo e admitindo alguns regimes especiais, a alíquota do novo IVA, se uniforme, ficaria no intervalo entre 20,73% e 22,02%.

IMB (2023) e Ministério da Fazenda (2023) também estimaram intervalos possíveis para a alíquota padrão neutra a partir de versões da PEC 45 que já previam isenções e alíquotas reduzidas, obtendo os seguintes intervalos, respectivamente: entre 27,3% e 30,7% e entre 25,45% e 27%. Porém o estudo cuja simulação capta de forma mais aproximada as diretrizes da reforma aprovada, e cuja metodologia é mais compatível com a nossa, é Afonso, Junior e Viana (2023), que estimou uma alíquota padrão neutra de 35%.

### 4.3 Impacto redistributivo das reformas

Os impactos redistributivos das reformas simuladas são mensurados tomando a despesa de consumo como indicador de bem-estar. Como o objetivo é comparar diferentes estruturas de alíquotas, esse indicador oferece estimativas mais significativas, pois permite identificar como a presença de alíquotas reduzidas e isenções imprime progressividade (ou regressividade) à tributação do consumo, afastando-a de um sistema perfeitamente proporcional. Antes de examinar os resultados, é útil observar a Tabela 7. Essa tabela agrupa os bens e serviços de acordo com a alíquota tributária que incide sobre eles no desenho da Reforma 3 e, em seguida, mostra a participação média de cada grupo na despesa monetária total dos domicílios, por quinto de despesa.[16]

**Tabela 7**: Participação dos itens agrupados por alíquota no orçamento das famílias (%)

| Nível da alíquota | Quintos de despesa total | | | | |
|---|---|---|---|---|---|
| | 1 | 2 | 3 | 4 | 5 |
| Zero | 14,7 | 11,8 | 10,3 | 8,4 | 4,4 |
| 40% da Alíquota padrão | 10,0 | 11,1 | 11,7 | 11,9 | 11,6 |
| 70% da Alíquota padrão | 0,3 | 0,4 | 0,8 | 1,0 | 2,0 |
| Alíquota padrão | 63,0 | 62,5 | 61,3 | 61,1 | 61,6 |
| Imposto Seletivo | 1,1 | 1,1 | 1,1 | 1,1 | 0,7 |
| Serviços financeiros e planos de saúde | 1,0 | 2,1 | 3,0 | 4,6 | 8,1 |
| **Total** | **100,0** | **100,0** | **100,0** | **100,0** | **100,0** |

Fonte: Cálculo dos autores a partir da POF 2017-2018.

Como esperado, o peso orçamentário do grupo de itens com alíquota zero, que consiste na cesta básica de alimentos, cai fortemente à medida que a despesa total aumenta. Já no caso dos outros grupos de bens e serviços com alíquotas reduzidas, a participação no orçamento é maior para famílias situadas mais no topo da distribuição. Em particular, destacam-se o grupo com 70% da alíquota padrão, cuja participação média no orçamento do último quinto é quase sete vezes maior do que no primeiro quinto; e o grupo 'serviços financeiros e planos de saúde', cujo peso é oito vezes maior na cesta de consumo dos 20% mais ricos do que dos 20% mais pobres. Vale ainda destacar que em nossa simulação pouco mais de 60% do consumo das famílias é tributado na alíquota padrão.

---

[16] Nesta seção, os quintos são construídos ordenando-se os domicílios de acordo com a despesa total de consumo (incluindo a despesa não-monetária) domiciliar *per capita*.

O impacto das reformas simuladas sobre o montante médio de tributos indiretos pago por domicílio em cada quinto de despesa é apresentado na Tabela 8. O primeiro ponto a observar é que, apesar das reformas representarem estruturas tributárias bem diferenciadas entre si, o impacto de todas elas é modesto, refletindo as limitações da tributação indireta como instrumento redistributivo. Observe-se que nas duas reformas em que o efeito redistributivo é mais significativo se destaca a presença de *cashback* – de forma focalizada, caso da Reforma 2, ou como uma transferência universal, caso da Reforma 4.

**Tabela 8**: Impacto das reformas sobre o montante de imposto pago por quinto de despesa

|  | Domicílios ordenados pela despesa total domiciliar per capita | | | | |
|---|---|---|---|---|---|
|  | Quintos | | | | |
|  | 1 | 2 | 3 | 4 | 5 |
| **Sistema vigente** | | | | | |
| Tributação (R$/mês) | 197 | 330 | 457 | 652 | 1.389 |
| Despesa (monetária) total (R$/mês) | 852 | 1.492 | 2.116 | 3.124 | 7.202 |
| **Reforma 1** | | | | | |
| Variação na tributação (R$/mês) | -22 | -25 | -24 | -13 | 84 |
| Variação na tributação/despesa total | -0,026 | -0,017 | -0,011 | -0,004 | 0,012 |
| **Reforma 2** | | | | | |
| Variação na tributação (R$/mês) | -136 | -12 | -6 | 14 | 140 |
| Variação na tributação/despesa total | -0,160 | -0,008 | -0,003 | 0,005 | 0,019 |
| **Reforma 3** | | | | | |
| Variação na tributação (R$/mês) | -52 | -33 | -33 | -15 | 133 |
| Variação na tributação/despesa total | -0,061 | -0,022 | -0,015 | -0,005 | 0,018 |
| **Reforma 4** | | | | | |
| Variação na tributação (R$/mês) | -100 | -52 | -31 | 7 | 176 |
| Variação na tributação/despesa total | -0,117 | -0,035 | -0,014 | 0,002 | 0,024 |

Nota: Os valores monetários estão expressos em reais de 2018.
Fonte: Cálculo dos autores a partir da POF 2017-2018.

As estimativas para a Reforma 1 indicam que a substituição da estrutura corrente de tributos indiretos por um IVA uniforme sobre todos os bens e serviços teria um efeito quase neutro do ponto de vista distributivo, porém, levemente progressivo. Esse resultado corrobora o obtido por Orair e Gobetti (2019), que realiza uma simulação análoga. Como já mencionado, isso reflete, essencialmente, o fato de serviços, que têm um peso maior no orçamento das classes de renda mais alta, serem subtributados no sistema vigente, de forma que a uniformização das alíquotas ao mesmo tempo que aumenta a carga tributária sobre os mais ricos permite reduzir a tributação de itens importantes no orçamento dos mais pobres, como energia elétrica e gás.

Como esperado, a Reforma 2, que simula a versão original da PEC 45 e consiste em um IVA uniforme, mas agora combinado a um imposto seletivo e *cashback* integral do IVA para os pobres, resulta em um ganho mais significativo para os 20% mais pobres dos domicílios, equivalente, em média, a 16% de sua despesa monetária.

A Reforma 3, inspirada na EC 132/2023, também é progressiva, ou seja, tem um efeito equalizador sobre a distribuição da carga tributária, porém, menos do que a Reforma 2, uma vez que os ganhos são menos focalizados nos mais pobres. Nessa reforma a redução na tributação do primeiro quinto equivalente a 6,1% de sua despesa monetária.

Por fim, observe que a Reforma 4, que substitui a isenção da cesta básica, na estrutura tributária da Reforma 3, por uma transferência universal de montante fixo e com impacto orçamentário igual ao da isenção da cesta, é mais progressiva do que a Reforma 3, produzindo um ganho para os 20% mais pobres quase duas vezes maior, em média. A transferência universal estimada representa, em média, cerca de 10% da renda monetária dos domicílios entre os 20% mais pobres. Note-se que para alguns domicílios a soma das transferências recebidas é maior do que o montante pago em imposto sobre o consumo, ou seja, o imposto se torna negativo.

A Tabela 9 compara medidas de pobreza e de desigualdade estimadas para o Brasil no contexto de cada reforma simulada, com aquelas mensuradas para o sistema tributário vigente, com o indicador de bem-estar utilizado sendo a despesa total líquida de tributos indiretos. Os resultados corroboram os apresentados na Tabela 8, indicando que o impacto das reformas é progressivo, mas modesto, reduzindo marginalmente a pobreza e a desigualdade no consumo.

**Tabela 9**: Impacto das reformas sobre a pobreza e a desigualdade de renda

| Indicadores de pobreza e desigualdade | Sistema corrente | Reforma 1 | Reforma 2 | Reforma 3 | Reforma 4 |
|---|---|---|---|---|---|
| Proporção de pobres | 0,2715 | 0,2655 | 0,2683 | 0,2632 | 0,2561 |
| Hiato de pobreza | 0,0963 | 0,0928 | 0,0749 | 0,0883 | 0,0809 |
| Hiato quadrático | 0,0474 | 0,0453 | 0,0338 | 0,0422 | 0,0367 |
| Coeficiente de Gini | 0,4930 | 0,4875 | 0,4785 | 0,4837 | 0,4781 |

Fonte: Cálculo dos autores a partir da POF 2017-2018.

Comparando a Reforma 2 com a Reforma 1, é interessante notar que apesar da intensidade da pobreza (medida pelo hiato de pobreza e o hiato quadrático) e o coeficiente de Gini serem menores na Reforma 2, a proporção de pobres é ligeiramente superior. Isso chama a atenção para a necessidade de considerar já no processo de definição de um esquema de *cashback* como será seu financiamento. Em nossa simulação, o *cashback* é financiado pelo aumento do próprio IVA e do Imposto Seletivo, de forma que alguns domicílios que não receberam *cashback* por se encontrarem um pouco acima da linha de pobreza antes da reforma, com a reforma, acabaram sendo puxados para pobreza.

Apesar desse efeito indesejado sobre a proporção de pobres, a Reforma 2 também produz uma redução maior na intensidade da pobreza e na desigualdade do que a Reforma 3. Também corroborando os resultados da Tabela 8, os resultados para Reforma 4 indicam que o pagamento de um montante fixo de dinheiro para todos os indivíduos na população seria mais eficaz na redução da pobreza e da desigualdade do que a isenção da cesta básica de alimentos. O índice de desigualdade de Gini, por exemplo, é reduzido em 0,9 ponto percentual com a isenção da cesta e em 1,5 ponto percentual com a transferência universal.

Esse resultado evidencia que o gasto tributário envolvido na isenção da cesta básica é mal focalizado, apesar da cesta ter um peso bem maior no orçamento das classes de renda mais baixa, como visto nas Tabelas 3 e 7. Estimamos que, enquanto 26% desse gasto é apropriado pelos domicílios que se encontram no quinto de renda mais alta, pouco menos de 16% beneficia os 20% mais pobres.

Em resumo, os resultados sugerem que os custos em termos de aumento de complexidade, perda de transparência e elevação da alíquota padrão decorrentes da diferenciação de alíquota, como na Reforma 3, não podem ser justificados com base em argumentos de equidade fiscal. Vale lembrar que a Reforma 2 tem uma única alíquota de IVA, enquanto na Reforma 3 há cinco alíquotas (incluindo a diferenciada sobre serviços financeiros). Note-se ainda que a alíquota do IVA uniforme na Reforma 2 foi estimada em 26,7%, enquanto a alíquota padrão estimada para a Reforma 3 é 33,2%, uma diferença de 6,5 pontos percentuais.

Por fim, a Tabela 10 mostra a variação marginal na alíquota padrão da Reforma 3, resultante da eliminação de cada tratamento diferenciado e do *cashback*, um por vez.

**Tabela 10**: Impacto das alíquotas favorecidas, *cashback* e imposto seletivo sobre a alíquota padrão

|  |  | Variação em p.p. |
|---|---|---|
| Alíquota padrão na Reforma 3 | 33,2% |  |
| Sem isenção da cesta básica | 28,9% | -4,3 |
| Sem alíquota de 40% da padrão | 29,4% | -3,8 |
| Sem alíquota de 70% da padrão | 33,0% | -0,2 |
| Sem alíquota reduzida para serviços financeiros | 32,3% | -0,9 |
| Sem *cashback* | 33,0% | -0,2 |
| Sem Imposto seletivo | 33,9% | 0,7 |

Fonte: Cálculo dos autores a partir da POF 2017-2018.

Observa-se que a tributação da cesta básica permitiria reduzir a alíquota padrão em 4,3 pontos percentuais, enquanto a eliminação da alíquota de 40% da padrão reduziria a alíquota padrão em 3,8 pontos percentuais. Vale destacar também que o tratamento especial de serviços financeiros e planos de saúde – cujo consumo se concentra no último quinto (55% do total) – tem um efeito significativo sobre a alíquota padrão, próximo de 1 ponto percentual. Por sua vez, o impacto da alíquota de 70% da padrão é equivalente ao impacto do *cashback* do IVA incidente sobre energia elétrica e gás de cozinha para os pobres. Por fim, verifica-se que na ausência do imposto seletivo, a alíquota padrão seria 0,7 ponto percentual mais elevada.

## 5 Comentários Finais

Combinando estimativas das alíquotas tributárias efetivas que incidem sobre os bens e serviços consumidos pelas famílias, obtidas a partir da matriz de insumo-produto, com os microdados da Pesquisa de Orçamentos Familiares, este estudo estimou os efeitos distributivos do sistema de tributação indireta vigente e utilizou o mesmo arcabouço metodológico para simular os impactos de reformas inspiradas na PEC 45/2019, em sua versão inicial e na forma aprovada pelo Congresso Nacional.

Os resultados indicam que no sistema atual de tributos indiretos a distribuição da carga tributária entre as famílias, mesmo calculada em relação à despesa de consumo (em vez da renda), é levemente regressiva. Isso significa que, apesar das isenções e multiplicidade de alíquotas, muitas vezes defendidas com base em argumentos de equidade, o sistema vigente fracassa em atenuar a carga tributária sobre os mais pobres.

Por outro lado, as simulações das estruturas de alíquotas baseadas no texto inicial da PEC 45/2019 e na EC 132/2023 sugerem que ambas têm um efeito equalizador, embora modesto. Porém, a forma original da PEC 45 – com um IVA uniforme e *cashback* para as famílias de baixa renda – é preferível à reforma aprovada, do ponto de vista distributivo. Isso acontece porque na proposta aprovada, apesar da isenção da cesta básica beneficiar proporcionalmente mais as famílias de baixa renda, as demais alíquotas favorecidas, por incidirem predominantemente sobre serviços, beneficiam os mais ricos – repetindo, portanto, um dos aspectos mais marcantes da regressividade do sistema atual.

Além do mais, em termos absolutos, a isenção da cesta básica produz um benefício maior para as classes se renda mais alta. Nossas simulações evidenciam esse problema ao mostrar que a substituição da isenção da cesta básica de alimentos, no contexto da reforma aprovada, por uma transferência fixa para todos os indivíduos na população teria um efeito mais significativo sobre a desigualdade e a pobreza.

Além de apontar a baixa efetividade redistributiva da diferenciação das alíquotas na tributação de bens e serviços, os resultados desse estudo indicam que o impacto das isenções e alíquotas reduzidas sobre a alíquota necessária para manter a arrecadação atual dos tributos que serão extintos pela reforma é potencialmente elevado. Enquanto, a alíquota padrão estimada para a proposta inicial da PEC 45 foi 26,7%, na versão simulada da EC 132 a alíquota neutra elevou-se para 33,2%.

Mais uma vez, convém ressalvar que, com relação à EC 132/2023, os detalhes sobre quais itens exatamente estarão sujeitos a isenções e alíquotas reduzidas, bem como a abrangência do Imposto Seletivo, ainda serão definidos por lei complementar. Antes dessa definição uma avaliação mais acurada dos impactos da reforma e de suas alíquotas fica inviabilizada. De todo modo, as simulações realizadas neste estudo dão uma ideia dos efeitos redistributivos potenciais da reforma e dos custos, em termos de impacto sobre a alíquota padrão, das escolhas que ainda precisam ser feitas no processo de regulamentação da reforma.

Outra ressalva que deve ser feita é que as simulações foram realizadas sob a hipótese de que as famílias não mudam suas parcelas orçamentárias em resposta a mudanças nos tributos. No entanto, pode-se esperar que haverá mudanças nos padrões de consumo em favor dos produtos que se tornarem relativamente mais baratos em detrimento daqueles que se tornarem mais caros. Na medida em que esses ajustamentos ocorram, o impacto distributivo e a receita do novo imposto tendem a ser reduzidos. Essa é uma questão que precisa ser explorada em trabalhos futuros.